# Paradox of description for motion of a hydrodynamic discontinuity in a potential and incompressible flow

Maxim Zaytsev*, V'yacheslav Akkerman**

*Moscow, Russian Federation,
** West Virginia University, WV 26506-6106, Morgantown, USA

## Abstract

Hydrodynamic discontinuities in an external potential and incompressible flow are investigated. Using the reaction front as an example in a 2D stream, an overdetermined system of equations is obtained that describes its motion in terms of the surface itself. Assuming that the harmonic flux approaching discontinuity is additional smooth, these equations can be used to determine the motion of this discontinuity without taking into account the influence of the flow behind the front, as well as the entire external flow. It is well known that for vanishingly low viscosity, the integral relation on the boundary (Dirichlet, Neumann problems) connects the tangential and normal components of the velocity. Knowing one of them along the boundary of the discontinuity, one can determine the entire external flow. However, assuming the external flow is smooth, this will also be the case for all derivatives of velocity with respect to coordinates and time. Then a paradox arises, knowing the position of the discontinuity and the velocity data at a point on its surface, it is possible to determine the motion of this discontinuity without taking into account the influence of the flow behind the front, as well as the entire external flow. There is no physical explanation for this mechanism. It is possible that a boundary layer is formed in front of the front, where viscosity plays a significant role and Euler equations are violated. It is argued that the classical idea of the motion of hydrodynamic discontinuities in the smooth, potential and incompressible flow in the external region should be supplemented in this case.

**1.** One of the most important issues arising in solving many hydrodynamic problems is the description of the motion of discontinuities [1-3]. In particular, tangential discontinuity occurs in the description of hydrodynamic instabilities (Rayleigh – Taylor, Darrieus – Landau, Kelvin – Helmholtz), gravitational waves, and other phenomena [1–4]. There are various equations for describing the motion of certain discontinuities in various approximations [4-8]. The authors also proposed equations describing the motion of some discontinuities [9–11]. However, these equations have not yet been studied in detail numerically and analytically. When considering the motion of discontinuities, a potential flow in front of the discontinuities is usually assumed. This creates a mathematical feature that can fundamentally change the idea of the motion of discontinuities propagating in potential and incompressible flows. It is well known that with vanishingly low viscosity, the integral relation on the boundary (the Dirichlet, Neumann problems) connects the tangential and normal components of the velocity [12]. Knowing one of them along the boundary of the discontinuity, we can determine the entire external flow. However, assuming additional smoothness of the external flow, this will also be the case for all derivatives of velocity with respect to coordinates and time. There is a paradox. Then, when known the position of the discontinuity and the velocity data at a point on its surface, it is possible to determine the motion of this discontinuity without taking into account the influence of the flow behind the front, as well as the entire external flow.

In this paper, we propose a system of equations for describing the discontinuity in the particular case of the reaction front, where this paradox appears.

**2.** Consider a 2D hydrodynamic discontinuity propagating in the external region with a potential and incompressible flow of unit density. Without loss of generality, we assume that the reaction front propagates (Fig. 1). We consider the equations of hydrodynamics in the region "-" before the discontinuity, where a potential flow is present.

$$\frac{\partial \mathbf{u}}{\partial t} + (\mathbf{u}\nabla)\mathbf{u} + \nabla P = 0, \tag{1}$$

$$\boldsymbol{\omega} = \nabla \times \mathbf{u} = 0, \tag{2}$$

$$\mathrm{div}\mathbf{u} = 0. \tag{3}$$

We assume that the fluid velocity and its derivatives with respect to time and coordinates approach the boundary of the discontinuity quite smoothly. The fluid velocity in the region "-" before the front from (2), (3) has the form $\mathbf{u} = \nabla\varphi$, $\Delta\varphi = 0$, where $\varphi$ is the velocity potential. On the boundary, each point $M$ moves with the speed $-V\mathbf{n}$, where $\mathbf{n}$ is the internal normal to the surface. We assume that the velocity of the planar reaction front relative to the stationary fuel is equal to unity, i.e.

$$\partial\varphi/\partial n + V = 1 \tag{4}$$

and $\mathbf{u}_\tau = \partial \varphi / \partial \tau$, where $\tau$ is the unit tangential vector to the surface.

Consider the Laplace equation for the velocity potential $\varphi$ in the two-dimensional region "-" before the discontinuity $\varphi$

$$\frac{\partial^2 \varphi}{\partial x^2} + \frac{\partial^2 \varphi}{\partial y^2} = 0. \tag{5}$$

We differentiate (5) in time $t$. We get that

$$\frac{\partial^2}{\partial x^2}\left(\frac{\partial \varphi}{\partial t}\right) + \frac{\partial^2}{\partial y^2}\left(\frac{\partial \varphi}{\partial t}\right) = 0. \tag{6}$$

Then, assuming that the flow approaching the discontinuity is smooth enough with respect to the coordinates and the time, the following relations are satisfied along the boundary of this discontinuity (Fig. 1) [9-12]:

$$\pi \varphi(\mathbf{r},t) = \oint \left[ \frac{\partial \varphi(\mathbf{r}_s,t)}{\partial n_s} \ln|\mathbf{r}_s - \mathbf{r}| - \varphi(\mathbf{r}_s,t) \mathbf{n}_s \cdot \frac{\mathbf{r}_s - \mathbf{r}}{|\mathbf{r}_s - \mathbf{r}|^2} \right] dl(\mathbf{r}_s), \tag{7}$$

$$\pi \frac{\partial \varphi(\mathbf{r},t)}{\partial t} = \oint \left[ \frac{\partial^2 \varphi(\mathbf{r}_s,t)}{\partial n_s \partial t} \ln|\mathbf{r}_s - \mathbf{r}| - \frac{\partial \varphi(\mathbf{r}_s,t)}{\partial t} \mathbf{n}_s \cdot \frac{\mathbf{r}_s - \mathbf{r}}{|\mathbf{r}_s - \mathbf{r}|^2} \right] dl(\mathbf{r}_s). \tag{8}$$

It is known from differential geometry that [9]

$$\frac{d\mathbf{n}}{dt} \cdot \tau = \frac{\partial V}{\partial \tau}, \quad \frac{d\mathbf{n}}{dt} \cdot \mathbf{n} = 0, \quad \frac{d\tau}{dt} \cdot \tau = 0, \quad \frac{\partial \mathbf{n}}{\partial \tau} = -\frac{1}{R}\tau, \quad \frac{\partial \tau}{\partial \tau} = \frac{1}{R}\mathbf{n}, \tag{9}$$

where $R$ - radius of curvature at a point at the front.

Also, assuming sufficient smoothness of the flow in coordinates and time, we have that [9]

$$\frac{d\varphi}{dt} = \frac{\partial \varphi}{\partial t} - V \frac{\partial \varphi}{\partial n}, \tag{10}$$

$$\frac{d}{dt}\left(\frac{\partial \varphi}{\partial n}\right) = \frac{d}{dt}(\mathbf{n} \cdot \nabla \varphi) = \left(\mathbf{n} \cdot \frac{d\nabla \varphi}{dt}\right) + \left(\frac{d\mathbf{n}}{dt} \cdot \nabla \varphi\right) = \frac{\partial^2 \varphi}{\partial t \partial n} - V\frac{\partial^2 \varphi}{\partial n^2} + \frac{\partial V}{\partial \tau}\frac{\partial \varphi}{\partial \tau}, \tag{11}$$

where $\frac{d}{dt} = \frac{\partial}{\partial t} - V\frac{\partial}{\partial n}$ - time derivative of front points moving with speed $-V\mathbf{n}$. From formula (5), it follows that at the front

$$\frac{\partial^2 \varphi}{\partial n^2} + \frac{\partial^2 \varphi}{\partial \tau^2} = 0. \tag{12}$$

We also have, using (9),

$$\left(\frac{\partial \varphi}{\partial \tau}\right)' = \frac{\partial \varphi}{\partial \tau}, \tag{13}$$

$$\left(\frac{\partial^2 \varphi}{\partial \tau^2}\right)' = \frac{\partial}{\partial \tau}(\boldsymbol{\tau} \cdot \nabla \varphi) = \left(\boldsymbol{\tau} \cdot \frac{\partial \nabla \varphi}{\partial \tau}\right) + \left(\frac{\partial \boldsymbol{\tau}}{\partial \tau} \cdot \nabla \varphi\right) = \frac{\partial^2 \varphi}{\partial \tau^2} + \frac{1}{R}\frac{\partial \varphi}{\partial n}, \qquad (14)$$

$$\left(\frac{\partial^2 \varphi}{\partial n \partial \tau}\right)' = \frac{\partial}{\partial \tau}(\mathbf{n} \cdot \nabla \varphi) = \left(\mathbf{n} \cdot \frac{\partial \nabla \varphi}{\partial \tau}\right) + \left(\frac{\partial \mathbf{n}}{\partial \tau} \cdot \nabla \varphi\right) = \frac{\partial^2 \varphi}{\partial \tau \partial n} - \frac{1}{R}\frac{\partial \varphi}{\partial \tau}, \qquad (15)$$

$$\left(\frac{\partial^3 \varphi}{\partial \tau^3}\right)' = \frac{\partial}{\partial \tau}\left(\frac{\partial^2 \varphi}{\partial \tau^2}\right)' = \left(\frac{\partial \boldsymbol{\tau}}{\partial \tau} \cdot \frac{\partial \nabla \varphi}{\partial \tau}\right) + \left(\boldsymbol{\tau} \cdot \frac{\partial^2 \nabla \varphi}{\partial \tau^2}\right) + \left(\frac{\partial^2 \boldsymbol{\tau}}{\partial \tau^2} \cdot \nabla \varphi\right) + \left(\frac{\partial \boldsymbol{\tau}}{\partial \tau} \cdot \frac{\partial \nabla \varphi}{\partial \tau}\right) =$$

$$= \frac{\partial^3 \varphi}{\partial \tau^3} + \frac{2}{R}\frac{\partial^2 \varphi}{\partial \tau \partial n} + \frac{\partial}{\partial \tau}\left(\frac{1}{R}\right)\frac{\partial \varphi}{\partial n} - \frac{1}{R^2}\frac{\partial \varphi}{\partial \tau}, \qquad (16)$$

$$\left(\frac{\partial^3 \varphi}{\partial n \partial \tau^2}\right)' = \frac{\partial}{\partial \tau}\left(\frac{\partial^2 \varphi}{\partial n \partial \tau}\right)' = \left(\frac{\partial \mathbf{n}}{\partial \tau} \cdot \frac{\partial \nabla \varphi}{\partial \tau}\right) + \left(\mathbf{n} \cdot \frac{\partial^2 \nabla \varphi}{\partial \tau^2}\right) + \left(\frac{\partial^2 \mathbf{n}}{\partial \tau^2} \cdot \nabla \varphi\right) + \left(\frac{\partial \mathbf{n}}{\partial \tau} \cdot \frac{\partial \nabla \varphi}{\partial \tau}\right) =$$

$$= -\frac{2}{R}\frac{\partial^2 \varphi}{\partial \tau^2} + \frac{\partial^3 \varphi}{\partial n \partial \tau^2} - \frac{\partial}{\partial \tau}\left(\frac{1}{R}\right)\frac{\partial \varphi}{\partial \tau} - \frac{1}{R^2}\frac{\partial \varphi}{\partial n}, \qquad (17)$$

…

where $(\ldots)'$ means the differentiation of values strictly along the boundary of the discontinuity.

Thus, according to formulas (10) - (14) the quantities $\partial \varphi(\mathbf{r},t)/\partial t$ and $\partial^2 \varphi(\mathbf{r},t)/\partial n \partial t$ in equation (8) are expressed in terms of $\partial \varphi/\partial n$, $\varphi$ and their time derivatives with respect to moving points on the discontinuity boundary and along the discontinuity boundary $\tau$ itself.

Let a moving curve, which is the boundary of the discontinuity, be defined as follows (Fig. 1): $x = x(t,\alpha)$, $y = y(t,\alpha)$, $\alpha \in [A,B]$. Every point $\alpha$, $\alpha \in [A,B]$ moves at a speed $-V\mathbf{n}$:

$$\frac{\partial \mathbf{r}}{\partial t} = -V\mathbf{n}, \quad \mathbf{r} = \mathbf{r}(t,\alpha) = \begin{pmatrix} x(t,\alpha) \\ y(t,\alpha) \end{pmatrix}. \qquad (18)$$

From differential geometry, it follows that

$$\frac{\partial s}{\partial \alpha} = \sqrt{\left(\frac{\partial x}{\partial \alpha}\right)^2 + \left(\frac{\partial y}{\partial \alpha}\right)^2}, \quad \frac{\frac{\partial \mathbf{r}}{\partial \alpha}}{\left|\frac{\partial \mathbf{r}}{\partial \alpha}\right|} = \boldsymbol{\tau}, \qquad (19)$$

where $s$ - the length of the curve, measured from a fixed point on this curve. Given (9), we have

$$\frac{\partial^2 \mathbf{r}}{\partial t \partial \alpha} = \frac{\partial(-V\mathbf{n})}{\partial \alpha} = -\frac{\partial V}{\partial \alpha}\mathbf{n} - V\frac{\partial \mathbf{n}}{\partial \alpha} = -\frac{\partial s}{\partial \alpha}\frac{\partial V}{\partial \tau}\mathbf{n} + \frac{\partial s}{\partial \alpha}\frac{V}{R}\boldsymbol{\tau}. \qquad (20)$$

Consequently,

$$\frac{\partial^2 s}{\partial \alpha \partial t} = \frac{\partial}{\partial t}\left(\sqrt{\left(\frac{\partial x}{\partial \alpha}\right)^2 + \left(\frac{\partial y}{\partial \alpha}\right)^2}\right) = \frac{\frac{\partial x}{\partial \alpha}\frac{\partial^2 x}{\partial t \partial \alpha} + \frac{\partial y}{\partial \alpha}\frac{\partial^2 y}{\partial t \partial \alpha}}{\sqrt{\left(\frac{\partial x}{\partial \alpha}\right)^2 + \left(\frac{\partial y}{\partial \alpha}\right)^2}} = \boldsymbol{\tau} \cdot \frac{\partial^2 \mathbf{r}}{\partial t \partial \alpha} = \frac{\partial s}{\partial \alpha}\frac{V}{R}. \tag{21}$$

Using this definition of a curve, we write equation (7) in the form [12]:

$$\pi \varphi(\mathbf{r}(t,\alpha),t) =$$

$$= \int_A^B \left[ \frac{\partial \varphi(\mathbf{r}_s(t,\alpha'),t)}{\partial n_s} \ln|\mathbf{r}_s(t,\alpha') - \mathbf{r}(t,\alpha)| - \varphi(\mathbf{r}_s(t,\alpha'),t)\mathbf{n}_s(t,\alpha') \cdot \frac{\mathbf{r}_s(t,\alpha') - \mathbf{r}(t,\alpha)}{|\mathbf{r}_s(t,\alpha') - \mathbf{r}(t,\alpha)|^2} \right] \frac{\partial s}{\partial \alpha'} d\alpha'$$

(22)

We differentiate (22) with respect to time $t$ using (9), (18), (21):

$$\pi \frac{d\varphi}{dt} = \int_A^B \left[ \frac{d}{dt}\left(\frac{\partial \varphi(\mathbf{r}_s,t)}{\partial n_s}\right) \ln|\mathbf{r}_s - \mathbf{r}| - \frac{d}{dt}(\varphi(\mathbf{r}_s,t))\mathbf{n}_s \cdot \frac{\mathbf{r}_s - \mathbf{r}}{|\mathbf{r}_s - \mathbf{r}|^2} \right] \frac{\partial s}{\partial \alpha'} d\alpha' +$$

$$+ \int_A^B \left[ \frac{\partial \varphi(\mathbf{r}_s,t)}{\partial n_s} \frac{(\mathbf{r}_s - \mathbf{r})(V\mathbf{n} - V_s\mathbf{n}_s)}{|\mathbf{r}_s - \mathbf{r}|^2} - \varphi(\mathbf{r}_s,t)\mathbf{n}_s \cdot \frac{(V\mathbf{n} - V_s\mathbf{n}_s)}{|\mathbf{r}_s - \mathbf{r}|^2} \right] \frac{\partial s}{\partial \alpha'} d\alpha' +$$

$$+ \int_A^B \left[ -\varphi(\mathbf{r}_s,t)\left(\frac{\partial V_s}{\partial \tau}\boldsymbol{\tau}_s \cdot \frac{\mathbf{r}_s - \mathbf{r}}{|\mathbf{r}_s - \mathbf{r}|^2} - 2\frac{[\mathbf{n}_s \cdot (\mathbf{r}_s - \mathbf{r})]}{|\mathbf{r}_s - \mathbf{r}|^4}[(\mathbf{r}_s - \mathbf{r}) \cdot (V\mathbf{n} - V_s\mathbf{n}_s)]\right) \right] \frac{\partial s}{\partial \alpha'} d\alpha' +$$

$$+ \int_A^B \left[ \frac{\partial \varphi(\mathbf{r}_s,t)}{\partial n_s} \ln|\mathbf{r}_s - \mathbf{r}| - \varphi(\mathbf{r}_s,t)\mathbf{n}_s \cdot \frac{\mathbf{r}_s - \mathbf{r}}{|\mathbf{r}_s - \mathbf{r}|^2} \right] \frac{\partial s}{\partial \alpha'} \frac{V_s}{R(\mathbf{r}_s,t)} d\alpha', \tag{23}$$

where $R$ - radius of curvature, $V = V(\mathbf{r}(t,\alpha),t)$ and $V_s = V(\mathbf{r}_s(t,\alpha'),t)$. We return to equation (23) to the integration over the curves. Then

$$\pi \frac{d\varphi}{dt} = \oint \left[ \frac{d}{dt}\left(\frac{\partial \varphi(\mathbf{r}_s,t)}{\partial n_s}\right) \ln|\mathbf{r}_s - \mathbf{r}| - \frac{d}{dt}(\varphi(\mathbf{r}_s,t))\mathbf{n}_s \cdot \frac{\mathbf{r}_s - \mathbf{r}}{|\mathbf{r}_s - \mathbf{r}|^2} \right] dl(\mathbf{r}_s) +$$

$$+ \oint \left[ \frac{\partial \varphi(\mathbf{r}_s,t)}{\partial n_s} \frac{(\mathbf{r}_s - \mathbf{r})(V\mathbf{n} - V_s\mathbf{n}_s)}{|\mathbf{r}_s - \mathbf{r}|^2} - \varphi(\mathbf{r}_s,t)\mathbf{n}_s \cdot \frac{(V\mathbf{n} - V_s\mathbf{n}_s)}{|\mathbf{r}_s - \mathbf{r}|^2} \right] dl(\mathbf{r}_s) +$$

$$+ \oint \left[ -\varphi(\mathbf{r}_s,t)\left(\frac{\partial V_s}{\partial \tau}\boldsymbol{\tau}_s \cdot \frac{\mathbf{r}_s - \mathbf{r}}{|\mathbf{r}_s - \mathbf{r}|^2} - 2\frac{[\mathbf{n}_s \cdot (\mathbf{r}_s - \mathbf{r})]}{|\mathbf{r}_s - \mathbf{r}|^4}[(\mathbf{r}_s - \mathbf{r}) \cdot (V\mathbf{n} - V_s\mathbf{n}_s)]\right) \right] dl(\mathbf{r}_s) +$$

$$+\oint\left[\frac{\partial\varphi(\mathbf{r}_s,t)}{\partial n_s}\ln|\mathbf{r}_s-\mathbf{r}|-\varphi(\mathbf{r}_s,t)\mathbf{n}_s\cdot\frac{\mathbf{r}_s-\mathbf{r}}{|\mathbf{r}_s-\mathbf{r}|^2}\right]\frac{V_s}{R(\mathbf{r}_s,t)}dl(\mathbf{r}_s). \tag{24}$$

Equation (24) is time-differentiated equation (7) differentiated by moving points on the discontinuity boundary. Given (4), (10) - (15), we obtain two different equations (8), (24) for definition $\partial\varphi/\partial n$ and $\varphi$ at the front, for which boundary conditions still need to be specified. In fact, derivatives $\frac{d}{dt}(\partial\varphi/\partial n)$ and $\frac{d\varphi}{dt}$ can be excluded from them. Then, using the formula (4), the front velocity $-V\mathbf{n}$ is determined.

More equations are obtained by differentiation with respect to coordinates. Let us once again consider the Laplace equation for the velocity potential $\varphi$ (5) in the two-dimensional region before the discontinuity. The values $\partial\varphi/\partial n$ and $\varphi$ at the boundary of the discontinuity are interconnected by Green's formula (7) [12]. In the region before the discontinuity, relations like (6) are also satisfied

$$\frac{\partial^2}{\partial x^2}\left(\frac{\partial\varphi}{\partial x}\right)+\frac{\partial^2}{\partial y^2}\left(\frac{\partial\varphi}{\partial x}\right)=0, \tag{25}$$

$$\frac{\partial^2}{\partial x^2}\left(\frac{\partial\varphi}{\partial y}\right)+\frac{\partial^2}{\partial y^2}\left(\frac{\partial\varphi}{\partial y}\right)=0. \tag{26}$$

For (25), (26), one can also write Green's formulas on the discontinuity boundary:

$$\pi\frac{\partial\varphi}{\partial x}(\mathbf{r})=\oint\left[\frac{\partial\left(\frac{\partial\varphi}{\partial x}\right)(\mathbf{r}_s)}{\partial n_s}\ln|\mathbf{r}_s-\mathbf{r}|-\frac{\partial\varphi}{\partial x}(\mathbf{r}_s)\mathbf{n}_s\cdot\frac{\mathbf{r}_s-\mathbf{r}}{|\mathbf{r}_s-\mathbf{r}|^2}\right]dl(\mathbf{r}_s), \tag{27}$$

$$\pi\frac{\partial\varphi}{\partial y}(\mathbf{r})=\oint\left[\frac{\partial\left(\frac{\partial\varphi}{\partial y}\right)(\mathbf{r}_s)}{\partial n_s}\ln|\mathbf{r}_s-\mathbf{r}|-\frac{\partial\varphi}{\partial y}(\mathbf{r}_s)\mathbf{n}_s\cdot\frac{\mathbf{r}_s-\mathbf{r}}{|\mathbf{r}_s-\mathbf{r}|^2}\right]dl(\mathbf{r}_s). \tag{28}$$

We transform formulas (27), (28) using the obvious relations on the boundary:

$$\frac{\partial\varphi}{\partial x}=\left(n_x\frac{\partial\varphi}{\partial n}+\tau_x\frac{\partial\varphi}{\partial\tau}\right), \tag{29}$$

$$\frac{\partial\varphi}{\partial y}=\left(n_y\frac{\partial\varphi}{\partial n}+\tau_y\frac{\partial\varphi}{\partial\tau}\right). \tag{30}$$

After substituting (29) into (27), we have

$$\pi\left(n_x\frac{\partial\varphi}{\partial n}+\tau_x\frac{\partial\varphi}{\partial\tau}\right)(\mathbf{r})=$$

$$= \oint \left[ \frac{\partial \left( n_x \frac{\partial \varphi}{\partial n_s} + \tau_x \frac{\partial \varphi}{\partial \tau_s} \right)(\mathbf{r}_s)}{\partial n_s} \ln|\mathbf{r}_s - \mathbf{r}| - \left( n_x \frac{\partial \varphi}{\partial n_s} + \tau_x \frac{\partial \varphi}{\partial \tau_s} \right)(\mathbf{r}_s) \mathbf{n}_s \cdot \frac{\mathbf{r}_s - \mathbf{r}}{|\mathbf{r}_s - \mathbf{r}|^2} \right] dl(\mathbf{r}_s)$$

or

$$\pi \left( n_x \frac{\partial \varphi}{\partial n} + \tau_x \frac{\partial \varphi}{\partial \tau} \right)(\mathbf{r}) =$$

$$= \oint \left[ \left( n_x \frac{\partial^2 \varphi}{\partial n_s^2} + \tau_x \frac{\partial^2 \varphi}{\partial n_s \partial \tau_s} \right)(\mathbf{r}_s) \ln|\mathbf{r}_s - \mathbf{r}| - \left( n_x \frac{\partial \varphi}{\partial n_s} + \tau_x \frac{\partial \varphi}{\partial \tau_s} \right)(\mathbf{r}_s) \mathbf{n}_s \cdot \frac{\mathbf{r}_s - \mathbf{r}}{|\mathbf{r}_s - \mathbf{r}|^2} \right] dl(\mathbf{r}_s)$$

or, given (12),

$$\pi \left( n_x \frac{\partial \varphi}{\partial n} + \tau_x \frac{\partial \varphi}{\partial \tau} \right)(\mathbf{r}) =$$

$$= \oint \left[ \left( -n_x \frac{\partial^2 \varphi}{\partial \tau_s^2} + \tau_x \frac{\partial^2 \varphi}{\partial n_s \partial \tau_s} \right)(\mathbf{r}_s) \ln|\mathbf{r}_s - \mathbf{r}| - \left( n_x \frac{\partial \varphi}{\partial n_s} + \tau_x \frac{\partial \varphi}{\partial \tau_s} \right)(\mathbf{r}_s) \mathbf{n}_s \cdot \frac{\mathbf{r}_s - \mathbf{r}}{|\mathbf{r}_s - \mathbf{r}|^2} \right] dl(\mathbf{r}_s). \quad (31)$$

Similarly

$$\pi \left( n_y \frac{\partial \varphi}{\partial n} + \tau_y \frac{\partial \varphi}{\partial \tau} \right)(\mathbf{r}) =$$

$$= \oint \left[ \left( -n_y \frac{\partial^2 \varphi}{\partial \tau_s^2} + \tau_y \frac{\partial^2 \varphi}{\partial n_s \partial \tau_s} \right)(\mathbf{r}_s) \ln|\mathbf{r}_s - \mathbf{r}| - \left( n_y \frac{\partial \varphi}{\partial n_s} + \tau_y \frac{\partial \varphi}{\partial \tau_s} \right)(\mathbf{r}_s) \mathbf{n}_s \cdot \frac{\mathbf{r}_s - \mathbf{r}}{|\mathbf{r}_s - \mathbf{r}|^2} \right] dl(\mathbf{r}_s). \quad (32)$$

Thus, taking into account (13) - (15), we obtained three integral-differential equations (7), (31) and (32) at the boundary of the discontinuity region for two surface unknowns $\partial \varphi / \partial n$ and $\varphi$ by which we can determine the front velocity $-V\mathbf{n}$ from (4), and, consequently, the movement of the entire front. Under the assumption that the flow approaches infinitely smoothly to discontinuity, composing equations of type (25), (26), but for the highest derivatives and using (13) - (17), we obtain infinitely many integral-differential equations for two surface unknowns $\partial \varphi / \partial n$ and $\varphi$.

As it is known, the harmonic function is infinitely differentiable within the domain of definition [12]. Let us single out a certain region in the external flow so that it lies entirely outside the reaction front. Then also on its boundary one can get an infinite number of integral-differential relations for only two surface unknowns $\partial \varphi / \partial n$ and $\varphi$. It is possible that not all of these equations are independent. Some are complex consequences of others (see Appendix B).

**3.** In this paper, we, in general, have proposed a way to reduce the dimension in the Laplace equation using Green's formulas as additional equations (see Appendix A). This is necessary to simplify and verify the calculations of this equation in applications. Using the reaction front as an example in a 2D stream, an overdetermined system of equations is obtained that describes its motion in terms of the surface itself. Assuming additional smoothness of the harmonic flow approaching discontinuity, one can determine the motion of this discontinuity without taking into account the influence of the flow behind the front, as well as the entire external flow. There is no physical explanation for this mechanism of movement of hydrodynamic discontinuity. It is possible that a boundary layer is formed in front of the front, where viscosity plays a significant role and Euler equations are violated. Thus, the motion of hydrodynamic discontinuities in the smooth, potential and incompressible flow in the external region should be explained precisely.

## Appendix A. Integral relations based on Green's formula for other equations of mathematical physics

Consider the stationary Laplace equation (5) in a certain region in $\mathbb{R}^2$. Let some closed surface move in it with a speed $-V\mathbf{n}$ that bounds the domain $G$ (Fig. 2). Then, on the surface, equations (10), (11), where there is no partial derivative with respect to time, are satisfied, as well as integral relations (7) and (24). The system of equations (10), (11) without a partial time derivative is equivalent to the Laplace equation (5). We have a system of equations (10), (11) overdetermined by an additional relation (7). For such a system of equations, one can apply the dimensionality reduction method described in [13, 14]. Namely, if a certain condition is fulfilled (equality to zero of solutions of a certain system of integral equations [14]), it is possible to obtain an overdetermined system of integral-differential equations only on this moving surface. One of these equations is equation (24).

In fact, only Green's integral formula for the Laplace equation is used here. It is possible, by analogy, to obtain integral relations for the other equations with the aim of overriding them, in order to reduce the dimension in them [13, 14].

Consider in a two-dimensional domain the evolutionary diffusion equation of the form

$$\frac{\partial \varphi}{\partial t} = \frac{\partial^2 \varphi}{\partial x^2} + \frac{\partial^2 \varphi}{\partial y^2} = \Delta \varphi . \tag{33}$$

We introduce the function $\psi$

$$\varphi = \Delta \psi .$$

Hence,

$$\frac{\partial \varphi}{\partial t} = \Delta \frac{\partial \psi}{\partial t} . \tag{34}$$

Let us consider in the domain of definition a certain movable surface bounding the domain $G$ (Fig. 2). For any point on this surface, Green's formula is satisfied [12]:

$$\pi\varphi(\mathbf{r},t) = \oint\left[\frac{\partial\varphi(\mathbf{r}_s,t)}{\partial n_s}\ln|\mathbf{r}_s-\mathbf{r}| - \varphi(\mathbf{r}_s,t)\mathbf{n}_s\cdot\frac{\mathbf{r}_s-\mathbf{r}}{|\mathbf{r}_s-\mathbf{r}|^2}\right]dl(\mathbf{r}_s) - \int_G \Delta\varphi(\mathbf{r}_s,t)\ln|\mathbf{r}_s-\mathbf{r}|d^2\mathbf{r}_s. \qquad (35)$$

Similarly, on this surface

$$\pi\frac{\partial\psi}{\partial t}(\mathbf{r},t) = \oint\left[\frac{\partial^2\psi(\mathbf{r}_s,t)}{\partial t\partial n_s}\ln|\mathbf{r}_s-\mathbf{r}| - \frac{\partial\psi}{\partial t}(\mathbf{r}_s,t)\mathbf{n}_s\cdot\frac{\mathbf{r}_s-\mathbf{r}}{|\mathbf{r}_s-\mathbf{r}|^2}\right]dl(\mathbf{r}_s) - \int_G \Delta\frac{\partial\psi}{\partial t}(\mathbf{r}_s,t)\ln|\mathbf{r}_s-\mathbf{r}|d^2\mathbf{r}_s.$$

$$(36)$$

Using formulas (33) and (34), we find that

$$\pi\varphi(\mathbf{r},t) = \oint\left[\frac{\partial\varphi(\mathbf{r}_s,t)}{\partial n_s}\ln|\mathbf{r}_s-\mathbf{r}| - \varphi(\mathbf{r}_s,t)\mathbf{n}_s\cdot\frac{\mathbf{r}_s-\mathbf{r}}{|\mathbf{r}_s-\mathbf{r}|^2}\right]dl(\mathbf{r}_s) - \int_G \frac{\partial\varphi}{\partial t}(\mathbf{r}_s,t)\ln|\mathbf{r}_s-\mathbf{r}|d^2\mathbf{r}_s, \qquad (37)$$

$$\pi\frac{\partial\psi}{\partial t}(\mathbf{r},t) = \oint\left[\frac{\partial^2\psi(\mathbf{r}_s,t)}{\partial t\partial n_s}\ln|\mathbf{r}_s-\mathbf{r}| - \frac{\partial\psi}{\partial t}(\mathbf{r}_s,t)\mathbf{n}_s\cdot\frac{\mathbf{r}_s-\mathbf{r}}{|\mathbf{r}_s-\mathbf{r}|^2}\right]dl(\mathbf{r}_s) - \int_G \frac{\partial\varphi}{\partial t}(\mathbf{r}_s,t)\ln|\mathbf{r}_s-\mathbf{r}|d^2\mathbf{r}_s.$$

$$(38)$$

Comparing (37) and (38), we find an integral relation of the form:

$$\pi\varphi(\mathbf{r},t) = \oint\left[\frac{\partial\varphi(\mathbf{r}_s,t)}{\partial n_s}\ln|\mathbf{r}_s-\mathbf{r}| - \varphi(\mathbf{r}_s,t)\mathbf{n}_s\cdot\frac{\mathbf{r}_s-\mathbf{r}}{|\mathbf{r}_s-\mathbf{r}|^2}\right]dl(\mathbf{r}_s) +$$

$$+\pi\frac{\partial\psi}{\partial t}(\mathbf{r},t) - \oint\left[\frac{\partial^2\psi(\mathbf{r}_s,t)}{\partial t\partial n_s}\ln|\mathbf{r}_s-\mathbf{r}| - \frac{\partial\psi}{\partial t}(\mathbf{r}_s,t)\mathbf{n}_s\cdot\frac{\mathbf{r}_s-\mathbf{r}}{|\mathbf{r}_s-\mathbf{r}|^2}\right]dl(\mathbf{r}_s). \qquad (39)$$

Integral relation (39) can be considered as an additional constraint equation to the system of equations (33), (34). In fact, (39) is Green's formula for the equation $\Delta\left(\varphi-\frac{\partial\psi}{\partial t}\right)=0$ obtained from (33), (34). A similar method is used to find the integral relation for the wave equation of the form:

$$\frac{\partial^2\varphi}{\partial t^2} = \frac{\partial^2\varphi}{\partial x^2} + \frac{\partial^2\varphi}{\partial y^2} = \Delta\varphi. \qquad (40)$$

Consider the following system of equations for a two-dimensional incompressible stationary fluid [1]

$$u_x \frac{\partial u_x}{\partial x} + u_y \frac{\partial u_x}{\partial y} + \frac{\partial P}{\partial x} = 0, \tag{41}$$

$$u_x \frac{\partial u_y}{\partial x} + u_y \frac{\partial u_y}{\partial y} + \frac{\partial P}{\partial y} = 0, \tag{42}$$

$$\frac{\partial u_x}{\partial x} + \frac{\partial u_y}{\partial y} = 0. \tag{43}$$

These equations can be represented as:

$$u_x \frac{\partial u_x}{\partial x} + u_y \frac{\partial u_x}{\partial y} + u_x \left( \frac{\partial u_x}{\partial x} + \frac{\partial u_y}{\partial y} \right) + \frac{\partial P}{\partial x} = \frac{\partial \left( u_x^2 + P \right)}{\partial x} + \frac{\partial \left( u_x u_y \right)}{\partial y} = 0, \tag{44}$$

$$\frac{\partial \left( u_x u_y \right)}{\partial x} + \frac{\partial \left( u_y^2 + P \right)}{\partial y} = 0, \tag{45}$$

$$\frac{\partial u_x}{\partial x} + \frac{\partial u_y}{\partial y} = 0. \tag{46}$$

Let's introduce the functions

$$\Delta \psi_1 = u_x^2 + P, \ \Delta \psi_2 = u_x u_y, \ \Delta \psi_3 = u_y^2 + P, \tag{47}$$

where $\Delta = \frac{\partial^2}{\partial x^2} + \frac{\partial^2}{\partial y^2}$. Then equations (44), (45) can be represented in the form

$$\Delta \left( \frac{\partial \psi_1}{\partial x} + \frac{\partial \psi_2}{\partial y} \right) = 0, \tag{48}$$

$$\Delta \left( \frac{\partial \psi_2}{\partial x} + \frac{\partial \psi_3}{\partial y} \right) = 0. \tag{49}$$

We have obtained the system of equations (46) - (49). On any closed curve in the plane from (48), (49), one can write down two integral Green's formulas connecting $\psi_1$, $\psi_2$, $\psi_3$ and their derivatives.

Consider general equations in the form [14]

$$H_k \left( U_1^1, ... U_v^i, ... U_p^m, S_1, ... S_v, ... S_p, \mathbf{x} \right) = 0, \ i = 1...m, \ v = 1...p, \ k = 1...p. \tag{50}$$

$$\frac{\partial U_v^i}{\partial x_m} = \frac{\partial U_v^m}{\partial x_i}, \ i = 1...(m-1), \ v = 1...p, \tag{51}$$

$$\frac{\partial S_v}{\partial x_m} = U_v^m, \ v = 1...p, \quad (52)$$

where $\mathbf{x} = (x_1,...x_m)$. Consider the case $m = 3$ and $m = 4$. Let's introduce the functions

$$\Delta \psi_v^i = U_v^i, \ \Delta \varphi_v = S_v, \ i = 1...m, \ v = 1...p, \quad (53)$$

where $\Delta = \dfrac{\partial^2}{\partial x_1^2} + ... + \dfrac{\partial^2}{\partial x_{m-1}^2}$. It follows from equations (51), (52) that

$$\Delta\left(\frac{\partial \psi_v^i}{\partial x_m} - \frac{\partial \psi_v^m}{\partial x_i}\right) = 0, \ i = 1...(m-1), \ v = 1...p, \quad (54)$$

$$\Delta\left(\frac{\partial \varphi_v}{\partial x_m} - \psi_v^m\right) = 0, \ v = 1...p. \quad (55)$$

Thus, we have the system of equations (50), (53) - (55). On any closed surface in space $\mathbb{R}^{m-1}$ from (54), (55) one can write down the $mp$ integral Green formulas connecting $\psi_v^i$, $\varphi_v$, $i = 1...m$, $v = 1...p$ and their derivatives. The peculiarity of this redefinition is that the resulting equations of reduced dimension are integral and their solution is determined immediately in the entire domain of definition of the original system of equations.

Note that in formulas (53) one can simply take $\Delta = \dfrac{\partial^2}{\partial x_1^2} + \dfrac{\partial^2}{\partial x_2^2}$ and obtain from (54) and (55) additional relations on some curve in space $\mathbb{R}^2$. The rest of the variables $\mathbf{x} = (x_3,...x_m)$ are considered as parameters in the integral Green's formulas. Then theoretically nothing prevents us from reducing the dimension in equations (50), (53) - (55) and transforming them to a system of integral-differential equations on this curve [13, 14].

## Appendix B. Some integral identities of the Green formula type

Consider some closed domain $G \subset \mathbb{R}^2$ and a function $\varphi(\mathbf{r}) \in C^\infty(\bar{G})$ defined on it (Fig. 2). For any point inside the region $G$, Green's formula is [12]:

$$2\pi\varphi(\mathbf{r},t) = \oint\left[\frac{\partial \varphi(\mathbf{r}_s,t)}{\partial n_s}\ln|\mathbf{r}_s - \mathbf{r}| - \varphi(\mathbf{r}_s,t)\mathbf{n}_s \cdot \frac{\mathbf{r}_s - \mathbf{r}}{|\mathbf{r}_s - \mathbf{r}|^2}\right]dl(\mathbf{r}_s) - \int_G \Delta\varphi(\mathbf{r}_s,t)\ln|\mathbf{r}_s - \mathbf{r}|d^2\mathbf{r}_s. \quad (56)$$

Let's write this formula for the function $\partial\varphi(\mathbf{r})/\partial x$

$$2\pi\frac{\partial\varphi}{\partial x}(\mathbf{r},t)=\oint\left[\frac{\partial\frac{\partial\varphi}{\partial x}(\mathbf{r}_s,t)}{\partial n_s}\ln|\mathbf{r}_s-\mathbf{r}|-\frac{\partial\varphi}{\partial x}(\mathbf{r}_s,t)\mathbf{n}_s\cdot\frac{\mathbf{r}_s-\mathbf{r}}{|\mathbf{r}_s-\mathbf{r}|^2}\right]dl(\mathbf{r}_s)-\int_G\Delta\frac{\partial\varphi}{\partial x}(\mathbf{r}_s,t)\ln|\mathbf{r}_s-\mathbf{r}|d^2\mathbf{r}_s.$$

(57)

We have,

$$\int_G\Delta\frac{\partial\varphi}{\partial x}(\mathbf{r}_s,t)\ln|\mathbf{r}_s-\mathbf{r}|d^2\mathbf{r}_s=\int_G\frac{\partial}{\partial x_s}\left(\Delta\varphi(\mathbf{r}_s,t)\ln|\mathbf{r}_s-\mathbf{r}|\right)d^2\mathbf{r}_s-\int_G\Delta\varphi(\mathbf{r}_s,t)\frac{\partial\ln|\mathbf{r}_s-\mathbf{r}|}{\partial x_s}d^2\mathbf{r}_s=$$

$$=\oint n_{xs}\Delta\varphi(\mathbf{r}_s,t)\ln|\mathbf{r}_s-\mathbf{r}|dl(\mathbf{r}_s)+\int_G\Delta\varphi(\mathbf{r}_s,t)\frac{\partial\ln|\mathbf{r}_s-\mathbf{r}|}{\partial x}d^2\mathbf{r}_s=$$

$$=\oint n_{xs}\Delta\varphi(\mathbf{r}_s,t)\ln|\mathbf{r}_s-\mathbf{r}|dl(\mathbf{r}_s)+\frac{\partial}{\partial x}\int_G\Delta\varphi(\mathbf{r}_s,t)\ln|\mathbf{r}_s-\mathbf{r}|d^2\mathbf{r}_s.$$

(58)

Substitute (56) into (58). Then for any point inside the region $G$ we have

$$\int_G\Delta\frac{\partial\varphi}{\partial x}(\mathbf{r}_s,t)\ln|\mathbf{r}_s-\mathbf{r}|d^2\mathbf{r}_s=\oint n_{xs}\Delta\varphi(\mathbf{r}_s,t)\ln|\mathbf{r}_s-\mathbf{r}|dl(\mathbf{r}_s)+$$

$$+\frac{\partial}{\partial x}\oint\left[\frac{\partial\varphi(\mathbf{r}_s,t)}{\partial n_s}\ln|\mathbf{r}_s-\mathbf{r}|-\varphi(\mathbf{r}_s,t)\mathbf{n}_s\cdot\frac{\mathbf{r}_s-\mathbf{r}}{|\mathbf{r}_s-\mathbf{r}|^2}\right]dl(\mathbf{r}_s)-2\pi\frac{\partial\varphi}{\partial x}(\mathbf{r},t).$$

(59)

Substitute (59) into (57). Then, after transformations, we find that

$$0=\oint\left[\frac{\partial\frac{\partial\varphi}{\partial x}(\mathbf{r}_s,t)}{\partial n_s}\ln|\mathbf{r}_s-\mathbf{r}|-\frac{\partial\varphi}{\partial x}(\mathbf{r}_s,t)\mathbf{n}_s\cdot\frac{\mathbf{r}_s-\mathbf{r}}{|\mathbf{r}_s-\mathbf{r}|^2}\right]dl(\mathbf{r}_s)-\oint n_{xs}\Delta\varphi(\mathbf{r}_s,t)\ln|\mathbf{r}_s-\mathbf{r}|dl(\mathbf{r}_s)-$$

$$-\frac{\partial}{\partial x}\oint\left[\frac{\partial\varphi(\mathbf{r}_s,t)}{\partial n_s}\ln|\mathbf{r}_s-\mathbf{r}|-\varphi(\mathbf{r}_s,t)\mathbf{n}_s\cdot\frac{\mathbf{r}_s-\mathbf{r}}{|\mathbf{r}_s-\mathbf{r}|^2}\right]dl(\mathbf{r}_s)$$

or

$$0=\oint\left[\left(-n_x\frac{\partial^2\varphi}{\partial\tau_s^2}+\tau_x\frac{\partial^2\varphi}{\partial n_s\partial\tau_s}\right)\ln|\mathbf{r}_s-\mathbf{r}|-\left(n_x\frac{\partial\varphi}{\partial n_s}+\tau_x\frac{\partial\varphi}{\partial\tau_s}\right)(\mathbf{r}_s,t)\mathbf{n}_s\cdot\frac{\mathbf{r}_s-\mathbf{r}}{|\mathbf{r}_s-\mathbf{r}|^2}\right]dl(\mathbf{r}_s)-$$

$$-\frac{\partial}{\partial x}\oint\left[\frac{\partial\varphi(\mathbf{r}_s,t)}{\partial n_s}\ln|\mathbf{r}_s-\mathbf{r}|-\varphi(\mathbf{r}_s,t)\mathbf{n}_s\cdot\frac{\mathbf{r}_s-\mathbf{r}}{|\mathbf{r}_s-\mathbf{r}|^2}\right]dl(\mathbf{r}_s).$$

(60)

Relation (60) holds for any $\partial\varphi/\partial n$ and $\varphi$ defined on the surface of the region $G$, but for any points within this region, since differentiation under the integral sign in (60) is possible only for these points. In the case when the point lies on the boundary of the region $G$ in (60) considered in this article, a separate study should be carried out.

Obviously, in addition to (60), considering the other derivatives of $\varphi(\mathbf{r})$, this method can be used to obtain an infinite number of similar integral identities.

## Appendix C. On the structure of the hydrodynamic flow caused by the movement of the reaction front

Let us consider the motion of the reaction front propagating from the ignition point in an external, incompressible 2D flow (Fig. 1). Let's denote the area in front of the front $G_-$ and the area behind the front $G_+$. Let us denote the gas velocity in the region before the discontinuity $G_-$ as $\mathbf{u}_-$ and in the region behind the discontinuity $G_+$ as $\mathbf{u}_+$. Let the external flow be potential $\boldsymbol{\omega}_- = \nabla \times \mathbf{u}_- = 0$ and have the form $\mathbf{u} = \nabla\varphi$, $\Delta\varphi = 0$, where $\varphi$ is the velocity potential.

Let the velocity undergo a discontinuity when passing through the reaction front, i.e.

$$\mathbf{u}_+ - \mathbf{u}_- = (\Theta - 1)\mathbf{n}, \tag{61}$$

where $\Theta = \rho_-/\rho_+ > 1$ is the gas density jump across the discontinuity, $\rho_-$ is the gas density ahead of the front, $\rho_+$ is the gas density behind the front, $\mathbf{n}$ is the internal normal to the surface.

Let us assume that the velocity flows are such that

$$\mathbf{u}_- \in C^1(G_-) \cap C(\bar{G}_-),\ \mathbf{u}_+ \in C^2(\bar{G}_+). \tag{62}$$

Let $\boldsymbol{\omega} = \nabla \times \mathbf{u}$. Consider the velocity field

$$\mathbf{u}_v = \frac{1}{2\pi} \nabla_\mathbf{r} \times \int_{G_+} \boldsymbol{\omega}(\mathbf{r}_s, t) \ln|\mathbf{r}_s - \mathbf{r}|\, d^2\mathbf{r}_s. \tag{63}$$

By virtue of (62), one can use the Ostrogradsky-Gauss formula [12] in expression (63)

$$\mathbf{u}_v = \frac{1}{2\pi} \int_{G_+} \boldsymbol{\omega}(\mathbf{r}_s, t) \times \nabla_\mathbf{r}\left(\ln|\mathbf{r}_s - \mathbf{r}|\right) d^2\mathbf{r}_s = -\frac{1}{2\pi} \int_{G_+} \boldsymbol{\omega}(\mathbf{r}_s, t) \times \nabla_{\mathbf{r}_s}\left(\ln|\mathbf{r}_s - \mathbf{r}|\right) d^2\mathbf{r}_s =$$

$$= \frac{1}{2\pi} \oint \left(\boldsymbol{\omega}(\mathbf{r}_s, t) \times \mathbf{n}_s\right) \ln|\mathbf{r}_s - \mathbf{r}|\, dl(\mathbf{r}_s) + \frac{1}{2\pi} \int_{G_+} \left(\nabla_{\mathbf{r}_s} \times \boldsymbol{\omega}(\mathbf{r}_s, t)\right) \ln|\mathbf{r}_s - \mathbf{r}|\, d^2\mathbf{r}_s. \tag{64}$$

It follows from (64) and the properties of the simple layer potential and the bulk potential [12] that the flow $\mathbf{u}_v$ is continuous through the front and can be extended so that

$$\mathbf{u}_{v-} \in C^1(\bar{G}_-),\ \mathbf{u}_{v+} \in C^1(\bar{G}_+), \tag{65}$$

where $\mathbf{u}_{v-} = \mathbf{u}_v$ is in the area in front of the gap $G_-$, $\mathbf{u}_{v+} = \mathbf{u}_v$ is in the area behind the gap $G_+$. We have,

$$\boldsymbol{\omega}_v = \nabla_\mathbf{r} \times \mathbf{u}_v = \frac{1}{2\pi} \nabla_\mathbf{r} \times \nabla_\mathbf{r} \times \int_{G_+} \boldsymbol{\omega}(\mathbf{r}_s,t) \ln|\mathbf{r}_s - \mathbf{r}| d^2\mathbf{r}_s = \frac{1}{2\pi} \Delta_\mathbf{r} \left( \int_{G_+} \boldsymbol{\omega}(\mathbf{r}_s,t) \ln|\mathbf{r}_s - \mathbf{r}| d^2\mathbf{r}_s \right) -$$

$$- \frac{1}{2\pi} \nabla_\mathbf{r} \left( \nabla_\mathbf{r} \cdot \int_{G_+} \boldsymbol{\omega}(\mathbf{r}_s,t) \ln|\mathbf{r}_s - \mathbf{r}| d^2\mathbf{r}_s \right). \tag{66}$$

Due to the fact that (62) $\nabla \cdot \boldsymbol{\omega} = \nabla \cdot [\nabla \times \mathbf{u}] = 0$ is satisfied in the region $G_+$ and $(\boldsymbol{\omega} \cdot \mathbf{n}) = 0$ at the front, then, using the Ostrogradsky-Gauss formula, we have

$$- \frac{1}{2\pi} \nabla_\mathbf{r} \left( \nabla_\mathbf{r} \cdot \int_{G_+} \boldsymbol{\omega}(\mathbf{r}_s,t) \ln|\mathbf{r}_s - \mathbf{r}| d^2\mathbf{r}_s \right) = \frac{1}{2\pi} \nabla_\mathbf{r} \left( \int_{G_+} \boldsymbol{\omega}(\mathbf{r}_s,t) \cdot \nabla_{\mathbf{r}_s} (\ln|\mathbf{r}_s - \mathbf{r}|) d^2\mathbf{r}_s \right) =$$

$$= -\frac{1}{2\pi} \nabla_\mathbf{r} \oint (\boldsymbol{\omega}(\mathbf{r}_s,t) \cdot \mathbf{n}_s) \ln|\mathbf{r}_s - \mathbf{r}| dl(\mathbf{r}_s) - \frac{1}{2\pi} \nabla_\mathbf{r} \int_{G_+} (\nabla_{\mathbf{r}_s} \cdot \boldsymbol{\omega}(\mathbf{r}_s,t)) \ln|\mathbf{r}_s - \mathbf{r}| d^2\mathbf{r}_s = 0.$$

Hence,

$$\boldsymbol{\omega}_v(\mathbf{r},t) = \frac{1}{2\pi} \Delta_\mathbf{r} \left( \int_{G_+} \boldsymbol{\omega}(\mathbf{r}_s,t) \ln|\mathbf{r}_s - \mathbf{r}| d^2\mathbf{r}_s \right) = \begin{cases} 0, & \mathbf{r} \in G_- \\ \boldsymbol{\omega}(\mathbf{r},t), & \mathbf{r} \in G_+ \end{cases}. \tag{67}$$

Consider a vector field $\mathbf{u}_p = \mathbf{u} - \mathbf{u}_v$. It is potentially in the regions $G_-$ and $G_+$, because according to (67) $\nabla \times \mathbf{u}_p = \nabla \times \mathbf{u} - \nabla \times \mathbf{u}_v = 0$. In addition, taking into account (63) and the assumption of incompressibility of the considered flows before and behind the front, we have $\nabla \cdot \mathbf{u}_p = \nabla \cdot \mathbf{u} - \nabla \cdot \mathbf{u}_v = 0$. Therefore, $\mathbf{u}_p = \nabla \phi_p$, $\Delta \phi_p = 0$ in the domains $G_-$ and $G_+$. Considering (62) and (65), it should be

$$\phi_{p-} \in C^2(G_-) \cap C^1(\bar{G}_-) \text{ и } \phi_{p+} \in C^2(\bar{G}_+), \tag{68}$$

where $\phi_{p-} = \phi_p$ is in the area in front of the gap $G_-$, $\phi_{p+} = \phi_p$ is in the area behind the gap $G_+$. From (61) and (64) it follows that the tangential component of the flow $\mathbf{u}_p = \partial \phi_p / \partial \tau$ does not change when passing through the front. We see that from (68) $\phi_{p+} \in C^2(\bar{G}_+)$ it follows that

$$\phi_{p-} \in C^2(G_-) \cap C^1(\bar{G}_-) \text{ и } \phi_{p+} \in C^2(G_+) \cap C^1(\bar{G}_+). \tag{69}$$

It is shown in [6] that if (69) is true and there is a discontinuity (61), then the potential $\phi_p$ in the domains $G_-$ and $G_+$ is necessary the Frankel potential [15]

$$\phi_p = \frac{1}{2\pi} \oint (\Theta - 1) \ln|\mathbf{r}_s - \mathbf{r}| dl(\mathbf{r}_s). \tag{70}$$

The potential $\phi_p$ defined by formula (70) is the potential of a simple layer [12]. It follows from the properties of this potential that, in general, it is not twice continuously differentiable in the closure $\bar{G}_+$. Consequently, conditions (68) may be violated.

Thus, the following paradox arises: in the general case, the hydrodynamic flow under conditions (61), (62) does not exist at all. It is possible that along some surface in the region behind the front $\bar{G}_+$ a discontinuity is formed in the field of the velocity $\mathbf{u}_+$ and/or its derivatives.

We multiply both parts of formula (61) by a vector $\mathbf{n}$ and integrate formula (61) along the surface of the reaction front. We have

$$\oint (u_{n+} - u_{n-}) dS = \oint (\Theta - 1) dS \tag{71}$$

or

$$\oint u_{n+} dS - \oint u_{n-} dS = (\Theta - 1) S \tag{72}$$

or, given the Ostrogradsky-Gauss formula,

$$-\int_{G_+} \mathrm{div}\mathbf{u}\, dV - \oint u_{n-} dS = (\Theta - 1) S \tag{73}$$

or, using (3),

$$\oint u_{n-} dS = -(\Theta - 1) S, \tag{74}$$

where $S$ is the surface area of the reaction front.

We have an additional relation (74) on the front surface, which depends on $\Theta = \rho_- / \rho_+$ and takes into account the structure of the flow in the reaction products. It can be used together with equations (8), (24) to determine the boundary conditions for them. Expression (74) can also be differentiated with respect to time on the moving surface of the front. In this way, one can also obtain relations on the surface that are necessary for closing the system (8), (24). One can also use the fact that behind the front, there are also relations of the form:

$$\mathrm{div}\left(\frac{\partial \mathbf{u}}{\partial t}\right) = 0 \tag{75}$$

and obtain more relations on the surface, using the Euler equations behind the front.

In the classical formulation of the problem of the motion of the reaction front, the Neumann problem for the external flow is posed before the front, i.e. the solution for the fluid velocity potential is sought in the space of functions $C^2(G_-) \cap C^1(\bar{G}_-)$. Behind the front, the solution is also sought in some function spaces. In this article, the region behind the front is not considered, but the problem is posed in the region $G_-$, which is a narrowing of the Neumann problem, i.e. the solution is sought in the space of functions $C^2(\bar{G}_-)$. Because of this, a closed system of front equations is obtained. The Euler equations contain the second derivatives of the velocity potential. The question is whether the Euler equations are valid up to the boundary, or whether the derivatives of velocity and pressure have a discontinuity at the boundary of the discontinuity. In the first case, it turns out that it is possible to determine the entire flow in the region $G_-$ without solving the equations of hydrodynamics behind the front. Possibly, the flow behind the front can be taken into account via the boundary or additional conditions on its surface (see (74)). In the second case, it is necessary to additionally investigate the flow ahead of the front, taking into account all the processes ahead of the front. It is still unknown which mathematical problems should be set to describe the movement of hydrodynamic discontinuities (Neumann, Dirichlet, Cauchy, etc.),

Note that we have obtained $\mathbf{u} = \mathbf{u}_p + \mathbf{u}_v$, where $\mathbf{u}_p = \nabla \phi_p$ (see (70)), $\mathbf{u}_v = \nabla_\mathbf{r} \times \boldsymbol{\psi}$, $\boldsymbol{\psi} = \frac{1}{2\pi} \int_{G_+} \boldsymbol{\omega}(\mathbf{r}_s, t) \ln |\mathbf{r}_s - \mathbf{r}| d^2 \mathbf{r}_s$, $\Delta \boldsymbol{\psi} = 0$ in the region $G_-$. It can be shown from the properties of the simple layer potential and the bulk potential $\boldsymbol{\psi} \in C^2(\bar{G}_-)$. Therefore, as shown in this paper, the each component of the vector function $\boldsymbol{\psi}$ and all its spatial derivatives can be defined on the front surface. Then by formula (4) or $\partial \phi_p / \partial n + (\mathbf{n} \cdot [\nabla \times \boldsymbol{\psi}]) + V = 1$ one can find the speed of the reaction front. For this structure of the velocity field to exist, it suffices to require $\mathbf{u}_- \in C^2(G_-) \cap C(\bar{G}_-)$, $\mathbf{u}_+ \in C^2(G_+) \cap C(\bar{G}_+)$ and $\boldsymbol{\omega} = \nabla \times \mathbf{u}_+ \in C(\bar{G}_+)$.

Figures

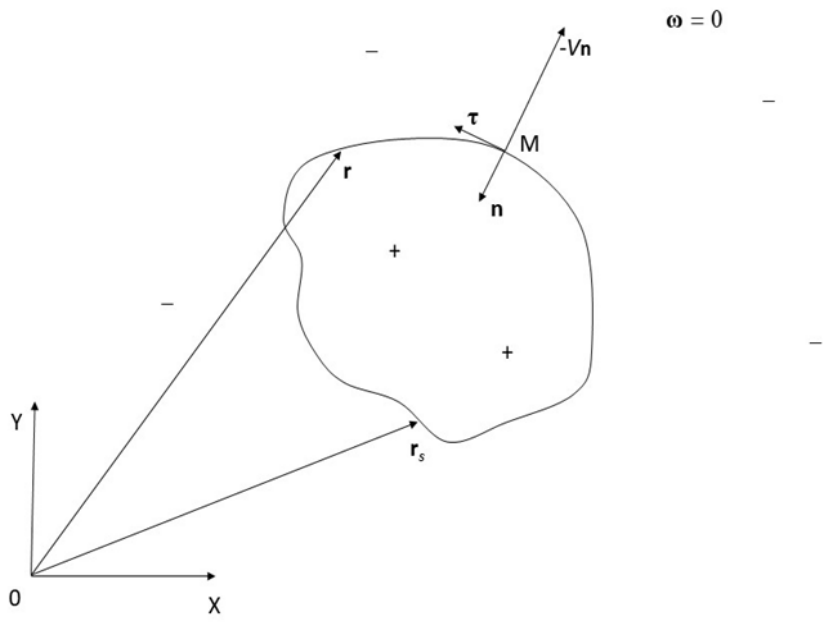

**Fig. 1.**

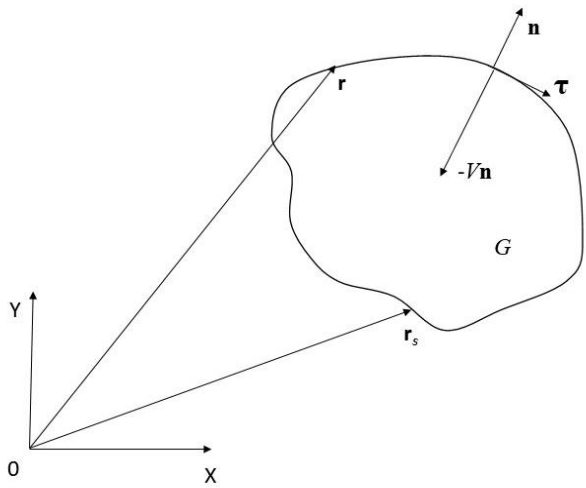

**Fig. 2.**

**Figure captions**

Fig. 1. Hydrodynamic discontinuity in an external potential 2D flow.
Fig. 2. Surface of a domain $G$.